\documentclass[11pt,draftcls,onecolumn]{IEEEtran}
\usepackage{amsmath,amssymb,bm,cite,nicefrac}
\usepackage[mathscr]{eucal}
\usepackage{upgreek}
\usepackage{fix-cm}
\newtheorem{thm}{Theorem}[section]

\newtheorem{lem}[thm]{Lemma}

\newtheorem{alg}[thm]{Algorithm}
\newtheorem{defn}{Definition}[section]

\newtheorem{rem}{Remark}[section]
\newcommand{\tc}{{\Breve\uptau}}

\DeclareMathOperator{\sgn}{sign}
\DeclareMathOperator{\Exp}{\mathbb{E}}
\DeclareMathOperator{\Prob}{\mathbb{P}}
\DeclareMathOperator*{\Argmin}{Arg\,min}
\newcommand{\D}{\mathrm{d}}
\newcommand{\E}{\mathrm{e}}
\newcommand{\df}{\,\triangleq\,}
\newcommand{\transp}{^{\mathsf{T}}}
\newcommand{\Uadm}{\mathfrak{U}}
\newcommand{\Usr}{\mathfrak{U}_{\mathrm{SM}}}
\newcommand{\Ussr}{\mathfrak{U}_{\mathrm{SSM}}}
\newcommand{\Lg}{L}
\newcommand{\sF}{\mathfrak{F}}

\newcommand{\Ind}{\mathbb{I}}
\newcommand{\Cc}{\mathcal{C}}
\newcommand{\Lp}{\mathcal{L}}

\newcommand{\Sob}{\mathscr{W}}
\newcommand{\Sobl}{\mathscr{W}_{\mathrm{loc}}}
\newcommand{\Act}{\mathbb{U}}
\newcommand{\imeas}{\mu}
\newcommand{\Lyap}{\mathcal{V}}
\newcommand{\RR}{\mathbb{R}}
\newcommand{\NN}{\mathbb{N}}
\newcommand{\fra}[2]{\leavevmode\kern.1em
 \raise.3ex\hbox{$\scriptstyle #1$}\kern-.1em
 {\scriptstyle /}\kern-.1em
 \lower.4ex\hbox{$\scriptstyle #2$}}
\newcommand{\abs}[1]{\lvert#1\rvert}

\newcommand{\norm}[1]{\lVert#1\rVert}

\begin{document}
\title{\Large On the Non-Uniqueness of Solutions to the Average Cost HJB
for Controlled Diffusions with Near-Monotone Costs}

\author{\Large Ari~Arapostathis
\thanks{Ari Arapostathis is with the Department of Electrical and
Computer Engineering, The University of Texas at Austin, Austin, TX~78712}
\thanks{This research was supported in part by the Office of Naval Research
through the Electric Ship Research and Development Consortium.}}

\maketitle

\begin{abstract}
We present a theorem for verification of optimality of controlled diffusions
under the average cost criterion with near-monotone running cost, without invoking
any blanket stability assumptions.
The implications of this result to the policy iteration algorithm are also
discussed.
\end{abstract}


\section{Introduction}

The theory of ergodic control of diffusions under near-monotone costs,
in the absence of blanket stability assumptions, lacks a satisfactory
result for the verification of optimality.
This has to do with the non-uniqueness of solutions to the
associated Hamilton--Jacobi--Bellman equation (HJB).
The results available assert the uniqueness (up to a constant)
of a value function which is bounded below that solves the HJB provided
the optimal value of the average cost, which appears in the equation as
a parameter, is selected.
However the optimal value of the average cost is unknown and this
leads to a circularity.
In an effort to fill this gap we present a verification theorem which
to the best of our knowledge is new.

We are concerned with
controlled diffusion processes $X = \{X_{t},\;t\ge0\}$
taking values in the $d$-dimensional Euclidean space $\RR^{d}$, and
governed by the It\^o stochastic differential equation
\begin{equation}\label{E-sde}
\D{X}_{t} = b(X_{t},U_{t})\,\D{t} + \upsigma(X_{t})\,\D{W}_{t}\,.
\end{equation}
All random processes in \eqref{E-sde} live in a complete
probability space $(\Omega, \sF,\Prob)$.
The process $W$ is a $d$-dimensional standard Wiener process independent
of the initial condition $X_{0}$.
The control process $U$ takes values in a compact, metrizable set $\Act$, and
$U_{t}(\omega)$ is jointly measurable in
$(t, \omega)\in[0,\infty)\times\Omega$.
Moreover, it is \emph{non-anticipative}:
for $s < t$, $W_{t} - W_{s}$ is independent of
\begin{equation*}
\sF_{s} \df \text{the completion of~} \sigma\{X_{0}, U_{r}, W_{r},\; r\le s\}
\text{~relative to~} (\sF,\Prob)\,.
\end{equation*}
Such a process $U$ is called an \emph{admissible control},
and we let $\Uadm$ denote the set of all admissible controls.
We impose fairly standard assumptions on the drift $b$
and the diffusion matrix $\upsigma$
to guarantee existence and uniqueness of solutions to \eqref{E-sde}, namely
that the diffusion is locally non-degenerate and
that $b$ and $\upsigma$ have at most affine growth,
are continuous and locally Lipschitz in $x$ uniformly in $u\in\Act$.
For the precise statements of these assumptions see Section~\ref{S-notA}.

Let $c\colon \RR^{d} \times\Act\to\RR$ be a nonnegative continuous function
which is referred to as the \emph{running cost}.
We assume that the cost function $c\colon\RR^{d}\times\Act\to\RR_{+}$ is continuous
and locally Lipschitz in its first argument uniformly in $u\in\Act$.
As well known, the ergodic control problem, in its \emph{almost sure}
(or \emph{pathwise}) formulation,
seeks to a.s.\ minimize over all admissible $U\in\Uadm$ the quantity
\begin{equation}\label{E-ergcrit}
\limsup_{t\to\infty}\; \frac{1}{t}\int_{0}^{t} c(X_{s},U_{s})\,\D{s}\,.
\end{equation}
A weaker, \emph{average} formulation seeks to minimize
\begin{equation}\label{E-avgcrit}
\limsup_{t\to\infty}\;\frac{1}{t}\int_{0}^{t}
\Exp^{U}\bigl[c(X_{s},U_{s})\bigr]\,\D{s}\,.
\end{equation}
We let $\varrho^{*}$ denote the infimum of \eqref{E-avgcrit} over all
admissible controls.
We assume that $\varrho^{*}<\infty$.

An fairly general class of running cost functions arising in practice
for which the ergodic control
problem is well behaved are the \emph{near-monotone} ones.
Let $M^{*}\in\RR_{+}\cup\{\infty\}$ be defined by
\begin{equation}\label{E-M}
M^{*}\df\liminf_{\abs{x}\to\infty}\;\min_{u\in\Act}\; c(x,u)\,.
\end{equation}
The running cost function $c$ is called near-monotone if $\varrho^{*}<M^{*}$.
Note that `norm-like' functions $c$ are always near-monotone.
The advantage of this class of problems is that no blanket stability (ergodicity)
assumption is imposed.
Indeed, models of controlled diffusions enjoying a uniform geometric ergodicity
do not arise often in applications.
What we frequently encounter in practice is a running cost which has a structure
that penalizes unstable behavior and thus renders all stationary optimal
controls stable.
Such is the case for quadratic costs typically used in linear control models.
Throughout this paper we assume that the running cost is near-monotone.

Solutions to the ergodic control problem can be
constructed via the HJB equation
\begin{equation}\label{E-HJB}
\sum_{i,j=1}^{d} a^{ij}(x)\,\frac{\partial^{2} V}
{\partial{x_{i}}\partial{x_{j}}}(x) + H(x,\nabla V) = \varrho\,,
\end{equation}
with $\varrho=\varrho^{*}$ where $a =[a^{ij}]$ is the symmetric
matrix $\frac{1}{2}\upsigma\,\upsigma\transp$ and
\begin{equation}\label{E-H}
H(x,p)\df\min_{u}\; \bigl[b(x,u)\cdot p + r(x,u)\bigr]\,.
\end{equation}
The real-valued function $V$ is bounded below in $\RR^{d}$ and lives in
$\Cc^{2}(\RR^{d})$, the space of twice-continuous differentiable functions on
$\RR^{d}$.
The resulting characterization is that a stationary Markov control $v^{*}$
is optimal for the ergodic control problem if and only if it is an a.e.\ measurable
selector from the minimizer of \eqref{E-HJB}, i.e., if and only if it satisfies
\begin{equation*}
b(x,v^{*}(x))\cdot \nabla V(x) + r(x,v^{*}(x)) = H\bigl(x,\nabla V(x)\bigr)
\quad\text{a.e.\ in~} \RR^{d}\,.
\end{equation*}

\subsection{Non-Uniqueness of solutions to the HJB}\label{S-nonunique}
Obtaining solutions to \eqref{E-HJB} is further complicated by the fact that
$\varrho^{*}$ is unknown.
Even though there exists a unique (up to a constant) solution
$V\in\Cc^{2}(\RR^{d})$ which is bounded below when $\varrho=\varrho^{*}$,
the HJB equation admits in general many solutions for $\varrho\ne\varrho^{*}$
\cite{BenBork-86b}.
We next review the example in \cite[Section~3.8.1]{book}.
Consider the one-dimensional controlled diffusion
$\D{X}_{t} = U_{t} \,\D{t} +\D{W}_{t}$, with
$\Act=[-1,1]$ and running cost $c(x) = 1-\E^{-\abs{x}}$.
If we define
\begin{equation*}
\xi_{\varrho} \df \log\frac{3}{2} + \log (1-\varrho)\,,\quad
\varrho\in\left[\tfrac{1}{3}, 1\right)
\end{equation*}
and
\begin{equation*}
V_{\varrho}(x) \df 2\int_{-\infty}^{x} \E^{2\abs{y-\xi_{\varrho}}}\D{y}
\int_{-\infty}^{y}
\E^{-2\abs{z-\xi_{\varrho}}}\bigl(\varrho -c(z)\bigr)\,\D{z}\,,
\quad x\in\RR\,,
\end{equation*}
then direct computation shows that
\begin{equation*}
\tfrac{1}{2}V''_{\varrho}(x) - \abs{V'_{\varrho}(x)}+c(x) = \varrho
\qquad\forall \varrho\in\left[\tfrac{1}{3}, 1\right)\,,
\end{equation*}
and so the pair $(V_{\varrho},\varrho)$ satisfies the HJB
for any $\varrho\in\left[\tfrac{1}{3}, 1\right)$.

With this example in mind, the question we pose is the following:
given a solution pair $(V,\varrho)$ of the HJB how does one verify if
a control obtained from the minimizer is indeed optimal, or equivalently
whether $\varrho=\varrho^{*}$?
As far as we know, the existing theory lacks a satisfactory verification theorem.

\subsection{A verification theorem}

We start by comparing the Markov control obtained
from the minimizer of the HJB for the example in Section~\ref{S-nonunique}
to the value of $\varrho$.
A stationary Markov control corresponding to the solutions of this HJB
is $w_{\varrho}(x) = -\sgn (x-\xi_{\varrho})$.
The controlled process under $w_{\varrho}$ has invariant probability
density $\psi_{\varrho}(x) = \E^{-2\abs{x-\xi_{\varrho}}}$.
A simple computation shows that
\begin{equation}\label{E-example1}
\int_{-\infty}^{\infty} c(x) \psi_{\varrho}(x)\,\D{x} = \varrho -\tfrac{9}{8}
(1-\varrho)(3\varrho-1)
\end{equation}
for all $\varrho\in\left[\tfrac{1}{3}, 1\right)$.
Thus if $\varrho>\frac{1}{3}$, then $\varrho$ is not the average cost
for the controlled process under $w_{\varrho}$.
This motivates the following definition.

\begin{defn}\label{D-comp}
A solution pair $(V,\varrho)\in\Cc^{2}(\RR^{d})\times\RR_{+}$ of the HJB equation
\eqref{E-HJB} with $\varrho\in[0,M^{*})$ is said to be \emph{compatible}
if $V$ is bounded below in $\RR^{d}$, and for some measurable selector
$v:\RR^{d}\to\Act$ from the minimizer of \eqref{E-HJB} the associated
invariant probability measure $\imeas_{v}$ of the diffusion controlled
by $v$ satisfies
\begin{equation}\label{E-beta}
\beta(v)\df\int_{\RR^{d}}c(x,v(x))\,\imeas_{v}(\D{x})=\varrho\,.
\end{equation}
\end{defn}

\begin{rem}
Since $V$ in Definition~\ref{D-comp} is bounded below it follows
from the Foster--Lyapunov stability criteria
that every measurable selector from the minimizer of the HJB is a Markov
control under which the diffusion is positive recurrent and hence
it admits an invariant probability measure (see \eqref{E-Lyap} in
Section~\ref{S-notC}).
\end{rem}

The main result of this paper
is the following.

\begin{thm}\label{T-unique}
Provided $c$ is near-monotone and bounded in $\RR^{d}\times\Act$ then
there exists a unique compatible
solution pair $(V,\varrho)\in\Cc^{2}(\RR^{d})\times\RR_{+}$
of the HJB equation \eqref{E-HJB}, with $V$ satisfying $V(0)=0$.
Moreover $\varrho= \varrho^{*}$ and any measurable selector $v:\RR^{d}\to\Act$ 
from the minimizer of \eqref{E-HJB} is an optimal stationary Markov control.
\end{thm}

Theorem~\ref{T-unique} offers a satisfactory verification theorem.
First note that by Theorem~\ref{T-unique}
if a solution pair $(V,\varrho)$ is compatible then every measurable selector
$v$ from the minimizer of the HJB satisfies \eqref{E-beta}.
Therefore it suffices to verify \eqref{E-beta} for any such $v$.
Going back to the example in Section~\ref{S-nonunique} it is clear from
\eqref{E-example1} that the pair $(V_{\varrho},\varrho)$ is compatible
for $\varrho=\frac{1}{3}$.
This suffices to assert that $\varrho^{*}=\frac{1}{3}$.

The organization of the paper is as follows.
In Section~\ref{S-not} we introduce the notation used in the paper, we
provide a precise statement concerning the assumptions on the model data,
and we review some basic definitions and results for controlled diffusions.
Section~\ref{S-main} is devoted to the proof of the main result.
In Section~\ref{S-PIA} we discuss some implications of the results
for the policy iteration algorithm.
Concluding remarks are in Section~\ref{S-concl}.

\section{Notation, Assumptions and some basic definitions}\label{S-not}

The standard Euclidean norm in $\RR^{d}$ is denoted by $\abs{\,\cdot\,}$.
The set of non-negative real numbers is denoted by $\RR_{+}$,
$\NN$ stands for the set of natural numbers, and $\Ind$ denotes
the indicator function.
We denote by $\uptau(A)$ the \emph{first exit time} of the process
$\{X_{t}\}$ from the set $A\subset\RR^{d}$, defined by
\begin{equation*}
\uptau(A) \df \inf\;\{t>0 : X_{t}\not\in A\}\,.
\end{equation*}
The closure, the boundary and the complement
of a set $A\subset\RR^{d}$ are denoted
by $\overline{A}$, $\partial{A}$ and $A^{c}$, respectively.
The open ball of radius $R$ in $\RR^{d}$, centered at the origin,
is denoted by $B_{R}$, and we let $\uptau_{R}\df \uptau(B_{R})$,
and $\tc_{R}\df \uptau(B^{c}_{R})$.

The term \emph{domain} in $\RR^{d}$
refers to a nonempty, connected open subset of the Euclidean space $\RR^{d}$. 
For a domain $D\subset\RR^{d}$,
the space $\Cc^{k}(D)$ ($\Cc^{\infty}(D)$)
refers to the class of all real-valued functions on $D$ whose partial
derivatives up to order $k$ (of any order) exist and are continuous,
and $\Cc_{b}(D)$ denotes the set of all bounded continuous
real-valued functions on $D$.
Also the space $\Lp^{p}(D)$, $p\in[1,\infty)$, stands for the Banach space
of (equivalence classes) of measurable functions $f$ satisfying
$\int_{D} \abs{f(x)}^{p}\,\D{x}<\infty$, and $\Lp^{\infty}(D)$ is the
Banach space of functions that are essentially bounded in $D$.
The standard Sobolev space of functions on $D$ whose generalized
derivatives up to order $k$ are in $\Lp^{p}(D)$, equipped with its natural
norm, is denoted by $\Sob^{k,p}(D)$, $k\ge0$, $p\ge1$.

In general if $\mathcal{X}$ is a space of real-valued functions on $Q$,
$\mathcal{X}_{\mathrm{loc}}$ consists of all functions $f$ such that
$f\varphi\in\mathcal{X}$ for every $\varphi\in\Cc_{c}^{\infty}(Q)$,
the space of smooth functions on $Q$ with compact support.
In this manner we obtain for example the space $\Sobl^{2,p}(Q)$.

We adopt the notation
$\partial_{i}\df\tfrac{\partial~}{\partial{x}_{i}}$ and
$\partial_{ij}\df\tfrac{\partial^{2}~}{\partial{x}_{i}\partial{x}_{j}}$
for $i,j\in\NN$.
We often use the standard summation rule that
repeated subscripts and superscripts are summed from $1$ through $d$.
For example,
\begin{equation*}
a^{ij}\partial_{ij}\varphi
+ b^{i} \partial_{i}\varphi \df \sum_{i,j=1}^{d}a^{ij}
\frac{\partial^{2}\varphi}{\partial{x}_{i}\partial{x}_{j}}
+\sum_{i=1}^{d} b^{i} \frac{\partial\varphi}{\partial{x}_{i}}\,.
\end{equation*}

\subsection{Assumptions on the Data}\label{S-notA}
The drift
$b=\bigl[b^{1},\dotsc,b^{d}\bigr]\transp :\RR^{d} \times\Act\mapsto\RR^{d}$
the diffusion matrix
$\upsigma=\bigl[\upsigma^{ij}\bigr]:\RR^{d}\mapsto\RR^{d\times d}$
and the running cost $c:\RR^{d} \times\Act\mapsto\RR_{+}$
are continuous and satisfy the following growth, local Lipschitz and
local non-degeneracy properties:
For each $R>0$ there exists a constant $\kappa_{R}$ such that
for all $x,y\in B_{R}$ and $u\in\Act$ it holds that
\begin{gather*}
\abs{b(x,u) - b(y,u)} + \norm{\upsigma(x) - \upsigma(y)}
\le \kappa_{R}\abs{x-y}\,,\\[2pt]
\abs{c(x,u) - c(y,u)} \le \kappa_{R}\abs{x-y}\,,\\[2pt]
\det[a(x)]\ge \kappa^{-1}_{R}\,,
\end{gather*}
and
\begin{equation*}
\abs{b(x,u)}^{2}+ \norm{\upsigma(x)}^{2}\le \kappa_{1}
\bigl(1 + \abs{x}^{2}\bigr)\,,\quad \forall (x,u)\in\RR^{d}\times\Act\,,
\end{equation*}
where
$\norm{\upsigma}^{2}\df\mathrm{trace}\left(\upsigma\upsigma\transp\right)$
and `$\det$' denotes the determinant.

\subsection{Controlled extended generator}

In integral form, \eqref{E-sde} is written as
\begin{equation}\label{E2}
X_{t} = X_{0} + \int_{0}^{t} b(X_{s},U_{s})\,\D{s}
+ \int_{0}^{t} \upsigma(X_{s})\,\D{W}_{s}\,.
\end{equation}
The second term on the right hand side of \eqref{E2} is an It\^o
stochastic integral.
We say that a process $X=\{X_{t}(\omega)\}$ is a solution of \eqref{E-sde},
if it is $\sF_{t}$-adapted, continuous in $t$, defined for all
$\omega\in\Omega$ and $t\in[0,\infty)$, and satisfies \eqref{E2} for
all $t\in[0,\infty)$ at once a.s.

With $u\in\Act$ treated as a parameter, we define
the family of operators $\Lg^{u}:\Cc^{2}(\RR^{d})\mapsto\Cc(\RR^{d})$ by
\begin{equation*}\label{E-Lu}
\Lg^{u} f(x) = a^{ij}(x)\,\partial_{ij}f(x)
+ b^{i}(x,u)\,\partial_{i}f(x)\,,\quad u\in\Act\,.
\end{equation*}
We refer to $\Lg^{u}$ as the \emph{controlled extended generator} of the diffusion.
The HJB equation in \eqref{E-HJB} then takes the form
\begin{equation*}
\min_{u\in\Act}\; \bigl[ \Lg^{u} V(x) + c(x,u)\bigr] = \varrho\,,\quad
x\in\RR^{d}\,.
\end{equation*}

Of fundamental importance in the study of functionals of $X$ is
It\^o's formula.
For $f\in\Cc^{2}(\RR^{d})$ and with $\Lg^{u}$ as defined in \eqref{E-Lu},
\begin{equation}\label{E-Ito}
f(X_{t}) = f(X_{0}) + \int_{0}^{t}\Lg^{U_{s}} f(X_{s})\,\D{s}
+ M_{t}\,,\quad\text{a.s.},
\end{equation}
where
\begin{equation*}
M_{t} \df \int_{0}^{t}\bigl\langle\nabla f(X_{s}),
\upsigma(X_{s})\,\D{W}_{s}\bigr\rangle
\end{equation*}
is a local martingale.
In this paper we also use Krylov's extension of the It\^o formula
\cite[p.~122]{Krylov} which extends \eqref{E-Ito} to functions $f$ in the
Sobolev space $\Sobl^{2,p}(\RR^{d})$, for $p>d$.

\subsection{Markov controls}\label{S-notC}

Recall that a control is called \emph{stationary Markov} if
$U_{t} = v(X_{t})$ for a measurable map $v :\RR^{d}\mapsto \Act$.
Correspondingly, the equation 
\begin{equation}\label{E5}
X_{t} = x_{0} + \int_{0}^{t} b\bigl(X_{s},v(X_{s})\bigr)\,\D{s} +
\int_{0}^{t} \upsigma(X_{s})\,\D{W}_{s}
\end{equation}
is said to have a \emph{strong solution} if given a Wiener process
$(W_{t},\sF_{t})$
on a complete probability space $(\Omega,\sF,\Prob)$, there exists a process $X$
on $(\Omega,\sF,\Prob)$, with $X_{0}=x_{0}\in\RR^{d}$, which is continuous,
$\sF_{t}$-adapted, and satisfies \eqref{E5} for all $t$ at once, a.s.
A strong solution is called \emph{unique}, if any two such solutions $X$ and
$X'$ agree $\Prob$-a.s., when viewed as elements of
$\Cc\bigl([0,\infty),\RR^{d}\bigr)$.
It is well known that under our assumptions on the data,
for any stationary Markov control $v$,
\eqref{E5} has a unique strong solution \cite{Gyongy-96}.

Let $\Usr$ denote the set of stationary Markov controls.
Under $v\in\Usr$, the process $X$ is strong Markov,
and we denote its transition function by $P^{v}(t,x,\cdot)$.
It also follows from the work of \cite{Bogachev-01} that under
$v\in\Usr$, the transition probabilities of $X$
have densities which are locally H\"older continuous.
Thus $\Lg^{v}$ defined by
\begin{equation*}
\Lg^{v} f(x) = a^{ij}(x)\,\partial_{ij}f(x)
+ b^{i}(x,v(x))\,\partial_{i}f(x)
\end{equation*}
for $v\in\Usr$ and $f\in\Cc^{2}(\RR^{d})$, is the generator of a
strongly-continuous semigroup on $\Cc_{b}(\RR^{d})$, which is strong Feller.
We let $\Prob_{x}^{v}$ denote the probability measure and
$\Exp_{x}^{v}$ the expectation operator on the canonical space of the
process under the control $v\in\Usr$, conditioned on the
process $X$ starting from $x\in\RR^{d}$ at $t=0$.

Recall that control $v\in\Usr$ is called \emph{stable}
if the associated diffusion is positive recurrent.
We denote the set of such controls by $\Ussr$,
and let $\imeas_{v}$ denote the unique invariant probability
measure on $\RR^{d}$ for the diffusion under the control $v\in\Ussr$.
It is well known that $v\in\Ussr$ if and only if there exists an
inf-compact function $\Lyap\in\Cc^{2}(\RR^{d})$, a bounded domain
$D\subset\RR^{d}$, and a constant $\varepsilon>0$ satisfying
\begin{equation}\label{E-Lyap}
\Lg^{v}\Lyap(x) \le -\varepsilon \qquad\forall x\in D^{c}\,.
\end{equation}

\section{Proof of the main result}\label{S-main}
The ergodic control problem for near-monotone cost functions is characterized
by Theorem~\ref{T3.1} below which combines Theorems~3.4.7, 3.6.6 and 3.6.10,
and Lemmas~3.6.8 and 3.6.9 in \cite{book}.

We need the following definition:
For $v\in\Ussr$, $\varrho>0$ and $r>0$ define
\begin{equation}\label{E-Psi}
\Psi^{v}_{r}(x;\varrho) \df \Exp_{x}^{v} \left[\int_{0}^{\tc_{r}}
\bigl(c(X_{t},v(X_{t}))-\varrho\bigr)\,\D{t}\right]\,,
\quad x\in B^{c}_{r}\,,
\end{equation}
where as defined in Section~\ref{S-not} $\tc_{r}$
stands for $\uptau(B^{c}_{r})$.
Note that $\Psi^{v}_{r}(x;\varrho)$ is always finite if $\beta(v)<\infty$,
with $\beta$ as defined in \eqref{E-beta}.

\begin{thm}\label{T3.1}
There exists a unique solution $V^{*}\in\Cc^{2}(\RR^{d})$
to the HJB equation
\begin{equation*}
\min_{u\in\Act}\; \bigl[ \Lg^{u} V^{*}(x) + c(x,u)\bigr] = \varrho^{*}\,,\quad
x\in\RR^{d}\,.
\end{equation*}
that is bounded below in $\RR^{d}$ and satisfies $V^{*}(0)=0$.
Also, a control $v^{*}\in\Usr$ is optimal with respect to the criteria
\eqref{E-ergcrit} and \eqref{E-avgcrit} if and only if it satisfies
\begin{equation*}
b(x,v(x))\cdot \nabla V^{*}(x) + r(x,v(x)) =
H\bigl(x,\nabla V^{*}(x)\bigr)\quad\text{a.e.\ in~} \RR^{d}\,.
\end{equation*}
Moreover, we have
\begin{align*}
V^{*}(x) &= \limsup_{r\downarrow 0}\;\inf_{v\in\Ussr}\;
\Psi^{v}_{r}(x;\varrho^{*})\nonumber\\[5pt]
&=\Psi^{v^{*}}_{r}(x;\varrho^{*})
+\Exp_{x}^{v^{*}} \bigl[ V^{*}(X_{\tc_{r}})\bigr]
\end{align*}
for all $x\in\RR^{d}$ and $r>0$.
\end{thm}

It follows by \eqref{E-Lyap} and the near-monotone hypothesis
that the optimal control $v^{*}$ in Theorem~\ref{T3.1} is stable.

We need the following lemma.

\begin{lem}\label{L3.2}
Let $(V,\varrho)\in\Cc^{2}(\RR^{d})\times\RR_{+}$ be a compatible solution pair
to \eqref{E-HJB} and $v:\RR^{d}\to\Act$ a measurable selector
from the minimizer of \eqref{E-H}.
Then
\begin{equation}\label{EL3.2a}
V(x) = \Psi^{v}_{r}(x;\varrho)
+ \Exp_{x}^{v}\bigl[V(X_{\tc_{r}})\bigr]\,,
\quad \forall r>0\,,\quad \forall x\in B^{c}_{r}\,.
\end{equation}
\end{lem}

\begin{IEEEproof}
By Dynkin's formula for any $R>r>0$ we have
\begin{multline}\label{EL3.2b}
V(x) = \Exp_{x}^{v} \biggl[\int_{0}^{\tc_{r}\wedge\uptau_{R}}
\bigl(c(X_{t},v(X_{t}))-\varrho\bigr)\,\D{t} \\
+ V(X_{\tc_{r}})\,\Ind\{\tc_{r}<\uptau_{R}\}
+V(X_{\uptau_{R}})\,\Ind\{\tc_{r}\ge\uptau_{R}\}\biggr]
\end{multline}
Since $V$ is bounded below
\begin{equation}\label{EL3.2c}
\liminf_{R\to\infty}\;
\Exp_{x}^{v}\bigl[V(X_{\uptau_{R}})\,\Ind\{\tc_{r}\ge\uptau_{R}\}\bigr]\ge0
\quad \forall x\in\RR^{d}\,.
\end{equation}
Applying Fatou's lemma to \eqref{EL3.2b} and using \eqref{EL3.2c} we obtain
\begin{equation}\label{EL3.2d}
V(x)\ge \Psi^{v}_{r}(x;\varrho)
+ \Exp_{x}^{v}\bigl[V(X_{\tc_{r}})\bigr]\,,
\quad \forall r>0\,,\quad \forall x\in B^{c}_{r}\,.
\end{equation}
From \eqref{EL3.2d} we obtain that
\begin{equation}\label{EL3.2e}
V(x) \ge V(0) + \lim_{r\downarrow 0}\; \Psi^{v}_{r}(x,\varrho)\qquad
\forall x\in\RR^{d}\,.
\end{equation}
Since $v\in\Ussr$, then by Lemma~{3.7.8}~(ii) in \cite{book}
the function $\Psi^{v}_{0}$ defined by
\begin{equation}\label{EL3.2f}
\Psi^{v}_{0}(x) \df \lim_{r\downarrow 0}\; \Psi^{v}_{r}(x,\beta(v))
\qquad\forall x\in\RR^{d}\,.
\end{equation}
lives in $\Sobl^{2,p}(\RR^{d})$, for any $p>d$, and satisfies
\begin{equation}\label{EL3.2g}
\Lg^{v} \Psi^{v}_{0}(x) + c(x,v(x)) = \beta(v)\qquad \text{on~}\RR^{d}\,.
\end{equation}
Since $\varrho=\beta(v)$, from \eqref{EL3.2e}--\eqref{EL3.2f}
we obtain $V-\Psi^{v}_{0}\ge V(0)$, and also by 
by \eqref{EL3.2g} we have $\Lg^{v}(V-\Psi^{v}_{0}) = 0$.
Therefore by the strong maximum principle we obtain $V-V(0)=\Psi^{v}_{0}$.
Also by (3.7.50) in \cite{book}
\begin{equation*}
\Psi^{v}_{0}(x) = \Psi^{v}_{r}(x;\beta(v))
+ \Exp_{x}^{v}\bigl[\Psi^{v}_{0}(X_{\tc_{r}})\bigr]\,,
\quad \forall r>0\,,\quad \forall x\in B^{c}_{r}\,.
\end{equation*}
from which \eqref{EL3.2a} follows since $\Psi^{v}_{0}=V-V(0)$.
\end{IEEEproof}

\begin{rem}
It follows by Lemma~\ref{L3.2} and \cite[Corollary~3.7.3]{book} that if
$(V,\varrho)\in\Cc^{2}(\RR^{d})\times\RR_{+}$ is a compatible solution pair
of \eqref{E-HJB} then
\begin{equation}\label{E-compat2}
\frac{1}{t}\;\Exp_{x}^{v}\bigl[V(X_{t})\bigr] \xrightarrow[t\to\infty]{}0\,.
\end{equation}
The converse also holds.
Therefore \eqref{E-compat2} can be used in the place of \eqref{E-beta} to
verify optimality of a solution to the HJB.
\end{rem}

We continue with the proof of the main result.

\begin{IEEEproof}[Proof of Theorem~\ref{T-unique}]
Let $(\Hat{V},\Hat{\varrho})$ be a compatible solution pair to \eqref{E-HJB}
and $\Hat{v}:\RR^{d}\to\Act$ an associated measurable selector
from the minimizer of \eqref{E-H}.
For each $R>0$ define
\begin{align*}
b_{R}(x,u) &\df \begin{cases} b(x,u)&\text{if~~} \abs{x}<R\\[2pt]
b(x,\Hat{v}(x))& \text{if~~} \abs{x}\ge R\,,\end{cases}\\[5pt]
c_{R}(x,u) &\df \begin{cases} c(x,u)&\text{if~~} \abs{x}<R\\[2pt]
c(x,\Hat{v}(x))&\text{if~~} \abs{x}\ge R\,.\end{cases}
\end{align*}
Consider the following family of diffusions, parameterized by $R>0$,
given by
\begin{equation}\label{E-sdeR}
\D{X}_{t} = b_{R}(X_{t},U_{t})\,\D{t} + \upsigma(X_{t})\,\D{W}_{t}\,,
\end{equation}
with associated running costs $c_{R}(x,u)$.
For each $\alpha\in(0,1]$ the discounted optimal cost $V_{\alpha}^{R}$
defined by
\begin{equation*}
V_{\alpha}^{R}(x) \df \inf_{U\in\Uadm}\;\Exp_{x}^{U}\left[
\int_{0}^{\infty} \E^{-\alpha s}c_{R}(X_{s},U_{s})\,\D{s}\right]
\end{equation*}
relative to the controlled diffusion in \eqref{E-sdeR} lives in
$\Sobl^{2,p}(\RR^{d})$, for any $p>d$, and satisfies
\begin{equation*}
a^{ij}(x)\,\partial_{ij}V_{\alpha}^{R}(x)
+H_{R}(x,\nabla V_{\alpha}^{R})
=\alpha V_{\alpha}^{R}(x)\qquad \text{a.e.~ in~} \RR^{d}\,,
\end{equation*}
with
\begin{equation}\label{E-HR}
H_{R}(x,p)\df
\min_{u\in\Act}\;\left[b_{R}^{i}(x,u)\,p + c_{R}(x,u)\right]\,.
\end{equation}
Note that $V_{\alpha}^{R}$ may not live in $\Cc^{2}(\RR^{d})$, since
$b_{R}$ and $c_{R}$ are not necessarily continuous in $x$ for $\abs{x}>\RR$.
Nevertheless, the compactness of the embedding
$\Sob^{2,p}(B_{R})\hookrightarrow \Cc^{1,r}(\overline{B_{R}})$,
$r<1-\frac{d}{p}$, for $p>d$,
implies that $\nabla V_{\alpha}^{R}$ is H\"older continuous in $\overline{B_{R}}$.
This has two implications:
\begin{enumerate}
\item
There exists a measurable selector from the minimizer in
the definition of the Hamiltonian $H_{R}$.
\item
The restriction of $V_{\alpha}^{R}$ to $B_{R}$ is in $\Cc^{2}(B_{R})$.
\end{enumerate}

Fix $R>0$.
The running cost $c_{R}$ is clearly near monotone for the diffusion
in \eqref{E-sdeR}.
Therefore we may apply the standard theory in \cite[Section~3.6.2]{book}
to assert that $V_{\alpha}^{R}(\cdot)-V_{\alpha}^{R}(0)$
converges uniformly on compact sets in $\RR^{d}$ to some
$V^{R}\in\Sobl^{2,p}(\RR^{d})$, for any $p>d$,
while $\alpha V_{\alpha}^{R}(0)$ tends to some constant
$\varrho_{R}$ as $\alpha\downarrow 0$, and that the pair
$(V^{R},\varrho_{R})$ satisfies
\begin{equation*}
a^{ij}(x)\,\partial_{ij}V^{R}(x) +H_{R}(x,\nabla V^{R})
=\varrho_{R}\qquad \text{a.e.~ in~} \RR^{d}\,.
\end{equation*}
It is also the case that $V^{R}$ is bounded below and admits the following
stochastic representation:
for any measurable selector $v_{R}$ from the minimizer in \eqref{E-HR} we have
\begin{equation}\label{E-sr2}
V^{R}(x) = \Psi^{v_{R}}_{R}(x;\varrho_{R})
+ \Exp_{x}^{v_{R}}\bigl[V^{R}(X_{\tc_{R}})\bigr]\,,
\quad \forall x\in B^{c}_{r}\,.
\end{equation}
Also $\varrho_{R} = \beta(v_{R})$.
It is also clear that $\varrho_{R}\le \Hat{\varrho}$ for all $R\ge0$.
This is because $\alpha V_{\alpha}^{R}(0)\bigr|_{R=0}\to \Hat{\varrho}$
as $\alpha\downarrow0$, and $V_{\alpha}^{R}(0)$ is non-increasing in $R$.
Since $v_{R}$ agrees with $\Hat{v}$ on $B_{R}^{c}$
and $\varrho_{R}\le \Hat{\varrho}$, we obtain
\begin{equation*}
\Psi^{v_{R}}_{R}(x;\varrho_{R})=\Psi^{\Hat{v}}_{R}(x;\varrho_{R})
\ge \Psi^{\Hat{v}}_{R}(x;\Hat\varrho)\,,
\end{equation*}
which together with Lemma~\ref{L3.2} and \eqref{E-sr2} implies that
\begin{equation}\label{E-bound}
\Hat{V}(x) \le V^{R}(x) + \max_{\partial B_{R}}\; \Hat{V}
- \min_{\partial B_{R}}\;V^{R}\qquad\forall x\in B_{R}^{c}\,.
\end{equation}
Therefore since
\begin{equation*}
\frac{1}{t}\;\Exp_{x}^{v_{R}}\bigl[V^{R}(X_{t})\bigr]
\xrightarrow[t\to\infty]{}0
\end{equation*}
by \cite[Corollary~3.7.3]{book}, the bound in
\eqref{E-bound} shows that the same applies to $\Hat{V}$.
Applying Dynkin's formula to
\begin{equation*}
\Lg^{v_{R}}\Hat{V}(x) + c(x,v_{R}(x))\ge \Hat{\varrho}
\end{equation*}
and using the just established fact that
$\frac{1}{t}\,\Exp_{x}^{v_{R}}\bigl[\Hat{V}(X_{t})\bigr]
\xrightarrow[t\to\infty]{}0$ and the definition $\varrho_{R} = \beta(v_{R})$
we obtain $\varrho_{R} \ge \Hat{\varrho}$.
Therefore it must be the case that
\begin{equation}\label{E-equal}
\Hat{\varrho} = \varrho_{R}\qquad \forall R>0\,.
\end{equation}

Define the function $F_{R}:[0,1]\to\RR_{+}$ by
\begin{equation*}
F_{R}(\alpha)=\begin{cases}
\alpha V_{\alpha}^{R}(0)&\text{for~}\alpha\in(0,1]\\[2pt]
\varrho_{R}&\text{for~}\alpha=0\,.
\end{cases}
\end{equation*}
It is a simple matter to verify that $\alpha\mapsto V_{\alpha}^{R}(0)$
is continuous on $(0,1]$.
Therefore $F_{R}$ is continuous on $[0,1]$ for each fixed $R>0$.
It is also evident that $R\mapsto F_{R}$ is non-increasing.
Therefore by Dini's theorem $F_{R}$ converges uniformly on $[0,1]$ to
some non-negative function $F_{\infty}$ as $R\to\infty$, and as a result
$F_{\infty}$ is continuous on $[0,1]$.
It is a standard matter to show that for any $\alpha\in(0,1]$ the function
$V_{\alpha}^{R}$ converges uniformly on compact sets of $\RR^{d}$ as $R\to\infty$
to some $V_{\alpha}^{\infty}\in\Sobl^{2,p}(\RR^{d})$, for any $p>d$,
and that the limit is a solution of
\begin{equation}\label{E-HJBd}
\min_{u\in\Act}\;\bigl[\Lg^{u}V_{\alpha}^{\infty}(x) + c(x,u)\bigr]
=\alpha V_{\alpha}^{\infty}(x)\,,\qquad x\in\RR^{d}\,.
\end{equation}
By elliptic regularity $V_{\alpha}^{\infty}\in\Cc^{2}(\RR^{d})$.
Since $c$ is bounded, \eqref{E-HJBd} has a unique nonnegative solution
in $\Cc^{2}(\RR^{d})$ which admits the stochastic representation
\begin{equation*}
V_{\alpha}^{\infty}(x) \df \inf_{U\in\Uadm}\;\Exp_{x}^{U}\left[
\int_{0}^{\infty} \E^{-\alpha s}c(X_{s},U_{s})\,\D{s}\right]\,.
\end{equation*}
It is well known that
$\alpha V_{\alpha}^{\infty}(0)\to \varrho^{*}$ as $\alpha\downarrow0$
\cite[Theorem~3.6.6]{book}.
Since $F_{R}$ converges uniformly on $[0,1]$, we have
\begin{align}\label{E-conv}
\lim_{R\to\infty}\;\varrho_{R} &=
\lim_{R\to\infty}\;\lim_{\alpha\downarrow0}\; F_{R}(\alpha)\nonumber\\[3pt]
&= \lim_{\alpha\downarrow0}\;F_{\infty}(\alpha) \nonumber\\[3pt]
&= \lim_{\alpha\downarrow0}\;\alpha V_{\alpha}^{\infty}(0) \nonumber\\[3pt]
&=\varrho^{*}\,.
\end{align}
By \eqref{E-equal} and \eqref{E-conv} we obtain $\Hat{\varrho}=\varrho^{*}$.
\end{IEEEproof}

\section{A Remark on The Policy Iteration Algorithm}\label{S-PIA}

A good part of the difficulty in obtaining a solution to the
HJB equation lies in the fact that the optimal cost $\varrho^{*}$ is not known.
The policy iteration (PIA) provides an iterative procedure for
obtaining the HJB equation via iterations of linear equations.

Recall the definitions of $M^{*}$, $\beta$, $\Psi^{v}_{r}$ and $\Psi^{v}_{0}$
in \eqref{E-M}, \eqref{E-beta}, \eqref{E-Psi} and \eqref{EL3.2f}, respectively.
Under the near-monotone hypothesis, if $v\in\Ussr$ and $\beta(v)<M^{*}$
then $\Psi^{v}_{0}$ is the unique solution $V$ of the Poisson equation
\begin{equation*}
\Lg^{v}V(x) + c(x,v(x)) = \beta(v)\,,\quad x\in\RR^{d}
\end{equation*}
in $\Sobl^{2,p}(\RR^{d})$, $p>d$, which is bounded below
and satisfies $V(0)=0$.
Note also that \eqref{E-Lyap} implies that
any control $v$ satisfying $\varrho_{v}<M^{*}$ is stable.

We write the PIA in following form:

\begin{alg}[Policy Iteration]\label{PIA}
\begin{itemize}
\item[1)]
Initialization. 
Set $k=0$ and select any $v_{0}\in\Usr$
such that $\beta(v_{0}) < M^{*}$.
Set $V_{0}=\Psi^{v}_{0}$ and $\varrho_{0}=\beta(v_{0})$.
\item[2)] Policy improvement.
Select an arbitrary $v_{k+1}\in\Usr$ which satisfies
\begin{equation*}
v_{k+1}(x)\in\Argmin_{u\in\Act}\;
\bigl\{b^{i}(x,u)\,\partial_{i} V_{k}(x) + c(x,u)\bigr\}\,,
\quad x\in\RR^{d}\,.
\end{equation*}
\item[3)] Value determination.
Let $V_{k+1}=\Psi^{v_{k+1}}_{0}$ and $\varrho_{k+1}=\beta(v_{k+1})$.
If $\varrho_{k+1}=\varrho_{k}$ stop.
\end{itemize}
\end{alg}

It is well known and straightforward to show that, provided $c$ is near-monotone,
then over any iteration of the PIA $\varrho_{k}$ is a non-increasing sequence.
Also for some positive numbers $\alpha_{k}$ and $\gamma_{k}$, $k\ge0$, such that
$\alpha_{k}\downarrow1$ and $\gamma_{k}\downarrow0$ as $k\to\infty$ it holds
that (see \cite[Theorem~4.4]{Meyn})
\begin{equation*}
\alpha_{k+1} V_{k+1}(x) + \gamma_{k+1} \le \alpha_{k} V_{k} + \gamma_{k}\quad
\forall k\in\NN\,.
\end{equation*}
The near monotone hypothesis along with the fact that $\varrho_{k}$
is non-increasing imply that the density of the invariant probability
measure $\imeas_{v_{k}}$ is locally bounded away from zero uniformly
in $k\in\NN$.
Using the above properties one can show that $V_{k}$ converges uniformly on
compact sets of $\RR^{d}$ to some $\Hat{V}\in\Cc^{2}(\RR^{d})$ which together
with the constant $\Hat{\varrho}\df\lim_{k\to\infty}\;\varrho_{k}$ form
a solution pair for the HJB equation \eqref{E-HJB}
(see \cite[Lemma~2 and Corollary~1]{OneFest}).

Observe that at every iteration the PIA returns
a pair of the form
$(V_{k},\varrho_{k})=(\Psi^{v_{k}}_{0},\beta(v_{k}))$.
Hence a pair $(V,\varrho)\in\Cc^{2}(\RR^{d})\times\RR_{+}$
is an equilibrium of the PIA if and only if it is a compatible
solution pair to the HJB equation.
Therefore, if the running cost is bounded, then
by Theorem~\ref{T-unique} the only equilibrium of the PIA is the
optimal pair $(V^{*},\varrho^{*})$ in Theorem~\ref{T3.1}.
However this does not imply that the PIA always converges to the optimal
pair $(V^{*},\varrho^{*})$.
Because is does not preclude the possibility that the iterates
$(V_{k},\varrho_{k})$ may have a limit point
$(\Hat{V},\Hat{\varrho})$ which is not an equilibrium of the PIA.
Observe that the map $v\mapsto\imeas_{v}$ from $\Ussr$
under the topology of Markov controls (see \cite[Section~2.4]{book})
to the set of invariant probability measures under the Prohorov topology
is not in general continuous.
As a result if $\{v_{k}\}\subset\Ussr$ is a sequence which converges
under the topology of Markov controls but $\{\imeas_{v_{k}}\}$ is not
tight, we may obtain
\begin{equation*}
\lim_{k\to\infty}\;\beta(v_{k}) > \beta\Bigl(\lim_{k\to\infty}\;v_{k}\Bigr)\,.
\end{equation*}

It is interesting to note that if we allow
a transfinite number of iterations then, provided the running cost is bounded,
convergence to the optimal value can be obtained.
We give a brief description of this transfinite recursion in the next paragraph.
For a more sophisticated use of transfinite iterations in dynamic programming
we refer the reader to \cite{Maitra-92}.

The recursion on the ordinals is defined as follows:
We denote the algorithm
as $(V_{k+1},\varrho_{k+1}) = \mathcal{T}(V_{k},\varrho_{k})$.
Note that $\mathcal{T}$ is not really a map since the measurable
selector from the minimizer at each step is not unique, but we don't delve
into the formalism of inductive
definability because the recursion is quite intuitive, and also because
it is straightforward to demonstrate that it terminates at a countable ordinal.

Let $\omega_{1}$ denote the first uncountable ordinal.
Let $V_{0}=\Psi^{v}_{0}$ and $\varrho_{0}=\beta(v_{0})$, and
for every ordinal $\xi<\omega_{1}$ define 
$\bigl\{(V_{\xi},\varrho_{\xi})\,, \xi<\omega_{1}\bigr\}$ by
\begin{equation}\label{E-transf}
(V_{\xi},\varrho_{\xi}) = \mathcal{T}
\Bigl(\lim_{\eta<\xi}\;V_{\eta}\,,\;\lim_{\eta<\xi}\;\varrho_{\eta}\Bigr)\,.
\end{equation}
If $\xi$ is a limit ordinal, then since Algorithm~\ref{PIA} (which is defined
on $\NN$) converges, it follows that
\begin{equation*}
\bigl(\Hat{V}(\xi),\,\Hat{\varrho}(\xi)\bigr) \df
\Bigl(\lim_{\eta<\xi}\;V_{\eta}\,,\;\lim_{\eta<\xi}\;\varrho_{\eta}\Bigr)
\end{equation*}
is a solution of the HJB.
Suppose $\bigl(\Hat{V}(\xi),\,\Hat{\varrho}(\xi)\bigr)$ is not
a compatible pair, otherwise the recursion terminates.
Then by \eqref{E-transf} we obtain $V_{\xi}=\Hat{V}(\xi)$ and
$\varrho_{\xi}< \Hat{\varrho}(\xi)$ which imply that
$(V_{\xi},\varrho_{\xi})$ does not solve the HJB.
Therefore we must have $\varrho_{\xi+1}<\varrho_{\xi}$.
Set $\delta_{\xi}\df \varrho_{\xi+1}-\varrho_{\xi}$.
Since only a countable number of the $\delta_{\xi}$ can be positive it follows
that there exists $\xi^{*}<\omega_{1}$ such that $\delta_{\xi^{*}}=0$.
Therefore the recursion terminates at a countable ordinal.

\section{Concluding Remarks}\label{S-concl}

Theorem~\ref{T-unique} fills a gap in the theory of ergodic control
of diffusions under near-monotone costs, albeit under the assumption
of bounded running cost.
This assumption was only used to assert that
\eqref{E-HJBd} has a unique non-negative solution.
Therefore whenever this can be established for a particular problem
the hypothesis of bounded running costs can be waived.

There are also some standard situations when the problem can be mapped
to an equivalent problem with bounded costs.
Suppose that $c$ satisfies
\begin{equation*}
\sup_{x\in\RR^{d},u,u'\in\Act}\;\frac{c(x,u')}{c(x,u)}<\infty\,.
\end{equation*}
A particular case when this happens is of course when the running cost does not
depend on the control.
We leave it to the reader to verify that if we define
\begin{equation*}
g(x)\df 1+\min_{u\in\Act}\;c(x,u)\,,\quad
\Tilde{\upsigma}\df\frac{\upsigma}{g}\,,\quad \text{and}\quad
\Tilde{b}\df\frac{b}{g}\,,
\end{equation*}
then the controlled diffusion with data $\Tilde{b}$, $\Tilde{\upsigma}$
and running cost
\begin{equation*}
\Tilde{c} \df \frac{\varrho^{*}}{\min_{\RR^{d}}\; g}+\frac{1+c-\varrho^{*}}{g}
\end{equation*}
is an equivalent optimal control problem which satisfies the
assumptions of Theorem~\ref{T-unique}, and hence the conclusions of
this theorem apply to the original problem.


\end{document}